\documentclass[twocolumn,prb]{revtex4-1}
\usepackage{graphicx}
\usepackage{epsfig}
\usepackage{multirow}
\usepackage{epstopdf}
\usepackage{gensymb}
\usepackage{nicefrac}

\begin{document}

\title{Monolayer and bilayer pentacene on Cu(111)}

\author{J. A. Smerdon$^1$}
\author{M. Bode$^2$}
\author{N. P. Guisinger$^1$}
\author{J. R. Guest$^1$}\email{jrguest@anl.gov}
\affiliation{$^1$Center for Nanoscale Materials, Argonne National Laboratory, Argonne IL 60439, USA}
\affiliation{$^2$Physikalisches Institut, Experimentalle Physik II, Universit\"{a}t W\"{u}rzburg, Am Hubland, 97074 W\"{u}rzburg, Germany}

%pacs codes relate to STM in study of surface structure, organic semiconductors, surface structure of thin films and electronic structure of organic materials

\begin{abstract}

The morphology and electronic structure of pentacene (Pn) deposited on Cu(111) was studied using scanning tunneling microscopy (STM) and spectroscopy (STS).  Deposition of a multilayer followed by annealing to reduce coverage to a monolayer results in the formation of either of two unique phases: a 2D herringbone structure previously unobserved for any linear acene, or a `random-tiling' structure.  Coverage greater than a monolayer promotes the formation of a bilayer phase similar to that observed for Pn/Ag(111).  STS shows that the electronic structure of the first layer is strongly modified due to its proximity to the substrate while the second layer exhibits nearly bulk-like electronic structure.

\end{abstract}
\maketitle
\newpage

\section{Introduction}

The interest in molecular analogs to semiconductor electronics stems from several factors, such as the ever-looming limit to the ultimate feature size achievable using conventional lithography and the appeal of bottom-up fabrication techniques.  Of increasing interest in the current energy landscape are organic photovoltaics (OPVs) \cite{Tang86-apl}. Although these are not yet as efficient as contemporary silicon-based PVs, they have several desirable properties such as flexibility, low cost and greatly increased optical absorption.

The ultimate performance of a molecular device depends amongst other factors on the nature of the interfaces between materials, including the contacts between the active molecules and the substrate \cite{Shen04-cpc}.  The desire to understand and improve these contacts is a major part of the impetus behind forming and studying molecular layers on surfaces.  Additionally, the action of such devices as single-molecule rectifiers \cite{Aviram74-cpl} and molecular heterojunctions based on discrete donor and acceptor molecules depends strongly on the relative alignment of energy levels in the molecules, which may be strongly modified by the interaction of substrate electrons with molecular orbitals \cite{Ishii99-am}.

\begin{figure}
\begin{center}
\includegraphics[width=0.45\textwidth]{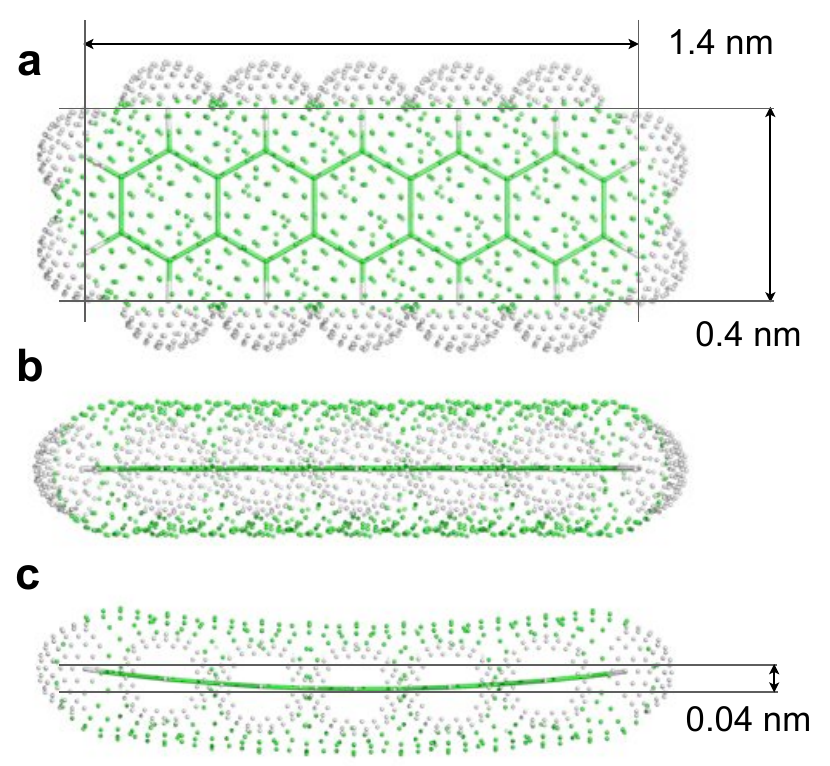}
\caption{\emph{(Color online)} \textbf{(a-b)}; Pentacene; \textbf{(c)};  Pentacene with a 0.04 nm curvature added as observed by M\"{u}ller \emph{et al.} \cite{Muller09-prb}.}
\label{pn}
\end{center}
\end{figure}

Pentacene (Pn) (C$_{22}$H$_{14}$), depicted in Figure \ref{pn}, is an archetypal $p$-type molecule for organic electronics, with good carrier mobility without doping \cite{Daraktchiev05-njp}. It is a linear polyacene comprising five benzene rings fused along C-C bonds.  In the bulk it adopts a herringbone structure \cite{Campbell61-ac}.  While deposition on inert surfaces such as SiO$_2$ or Bi/Si(111)-$7\times7$ leads to a similar but not identical structure with the molecular axes nearly perpendicular to the substrate surface \cite{Bouchoms99-sm,Sadowski05-apl}, deposition on metallic surfaces such as Au, Cu or Ag result in adsorption with the molecular plane parallel to  the surface.  On Au \cite{Schroeder02-jap,Kafer07-prb,France02-nl,Kang03-apl,Soe09-prl} and Ag \cite{Dougherty08-jpcc,Eremtchenko05-prb,Kafer07-cpl} the molecule appears flat; however, on Cu surfaces the molecule acquires a bending that is clear in scanning tunneling microscopy (STM) \cite{Muller09-prb,Annese07-ss,Lagoute04-prb,Lukas01-prl,Ferretti07-prl,Chen03-l} and atomic force microscopy (AFM) \cite{Gross09-science} measurements and has been quantified using DFT for adsorption on Cu(110) \cite{Muller09-prb}.  The bending on Cu(110) is approximately 0.04 nm measured between the center of the molecule and a plane bisecting the H atoms at either end and is shown in Figure \ref{pn} (c). Adsorption on Cu(111) has also been studied using DFT, though during relaxation the C atoms were constrained to be parallel to the surface, so molecular bending could not be reproduced \cite{Toyoda09-jes}.  There are no reports on nanoscale studies of the adsorption of Pn on a Cu(111) surface to monolayer coverage and beyond.

This is a report of the morphology and electronic structure of Pn on Cu(111) obtained through scanning tunneling microscopy/spectroscopy (STM/STS) measurements.

\begin{figure*}
\begin{center}
\includegraphics[width=0.9\textwidth]{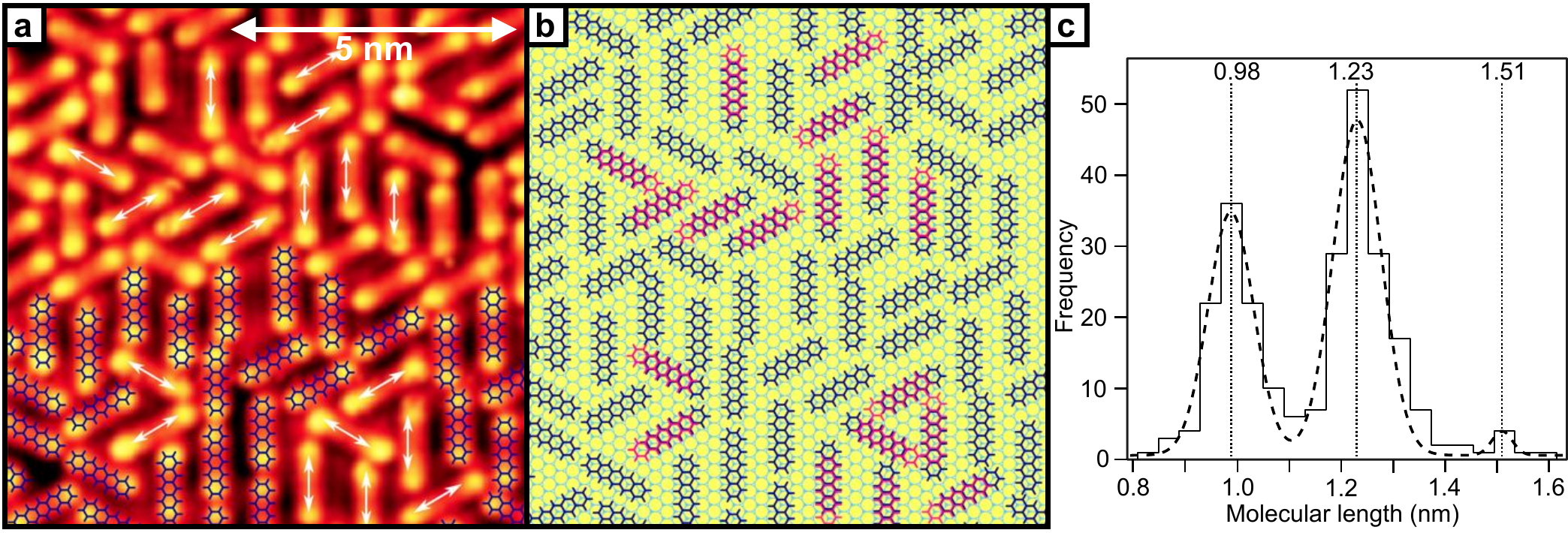}
\caption{\emph{(Color online)} \textbf{(a)}; 10 nm $\times$ 10 nm STM topograph ($V_{gap} = 0.5$ V, $I_T = 1$ nA) of Pn/Cu(111), showing the `random-tiling' structure with a section of the model proposed in \emph{(b)} superimposed, and with mobile molecules indicated with arrows showing the direction of motion; \textbf{(b)};  a model for a section of the Pn `random-tiling' structure, based on the adsorption site for a single molecule \cite{Lagoute04-prb}; \textbf{(c)}; histogram showing the distribution of apparent lengths of Pn molecules.}
\label{random}
\end{center}
\end{figure*}

\section{Experimental}

All measurements were carried out in a commercial Omicron UHV VT-AFM/STM operating with the sample maintained at 55 K.  The Cu(111) crystal was cleaned by simultaneously sputtering with Ar$^{+}$ ions at 1 keV and annealing to 900 K, with a final sputtering cycle at room temperature followed by annealing to 900 K.  Tungsten tips were prepared via electrochemical etching with 2 M NaOH solution and were cleaned with running water for 5 minutes and examined in an optical microscope before introduction to the UHV system.  

Pentacene was evaporated from a Dodecon 4-cell organic MBE source at 510 K, producing a multilayer that was then reduced to around a monolayer by post-deposition annealing as described in the text.  Each film described was prepared from the same starting point with a freshly cleaned Cu(111) surface.  All gap voltages are quoted as the sample bias.  Spectroscopy dI/dV results are collected with the standard lock-in technique ($f = 20$ kHz, $V_{mod}= 30$ mV).

During scanning, there was a tendency to collect molecules at the end of the tip, as evidenced by the appearance of multiple tips.  When this occurred, the tip was $e$-beam heated with 2 W for 10 s, which usually restored the imaging properties.

\section{Results and discussion}

\subsection{Scanning tunneling microscopy}

When Pn is deposited on the Cu(111) surface at low coverage, it has been demonstrated that it adopts a symmetrical adsorption site with the molecular axis lying along a close-packed $[1\bar{1}0]$ direction and with the molecule center above an hcp hollow site with a $b$-plane atom beneath \cite{Lagoute04-prb}.  There are no reports of monolayer coverages of Pn on Cu(111), though on Cu(110), Pn forms rows of Pn molecules with long axes parallel and molecular plane parallel to the surface \cite{Chen03-l,Lukas01-prl}.  On Cu(119), Pn forms long chains using the substrate steps as a template \cite{Gavioli05-prb,Annese07-ss}.  There are some reports of a freely-moving Pn `gas' at room temperature on Ag(111) \cite{Dougherty08-jpcc} and on a Pn monolayer adsorbed on Ag-Si(111)-($\sqrt{3}\times\sqrt{3}) R30^{\circ}$ \cite{Teng08-ss}.  Thermal motion in our experiments is expected to be suppressed because they were performed with a substrate temperature of 55 K.

\subsubsection{Pn/Cu(111) `random-tiling' - $R$-phase}

Following a post-deposition anneal to 420 K for 10 minutes, a structure emerges that we describe as a `random-tiling' of Pn molecules (shown in Fig. \ref{random} (a)). The bright ends of the Pn molecules are indicative of the aforementioned bending (indicated in Fig. \ref{pn}(c)).  Pentacene molecules are evenly distributed along the three main axes of the underlying Cu substrate, and knowledge of the adsorption site for a single molecule as determined using STM by Lagoute \emph{et al.} \cite{Lagoute04-prb} permits the identification of some of the relative positions of the molecules.  A structure model based on this adsorption preference is depicted in Figure \ref{random} (b), and shows reasonable agreement with the image.  The structure complexity and lack of simultaneous substrate resolution makes the exact determination of some nucleation sites intractable; nevertheless, a reasonable fit may be obtained, and various portions of the structure each yield a film density of $0.034 \pm 0.002$ Pn molecules per surface Cu atom.  

\begin{figure*}
\begin{center}
\includegraphics[width=0.9\textwidth]{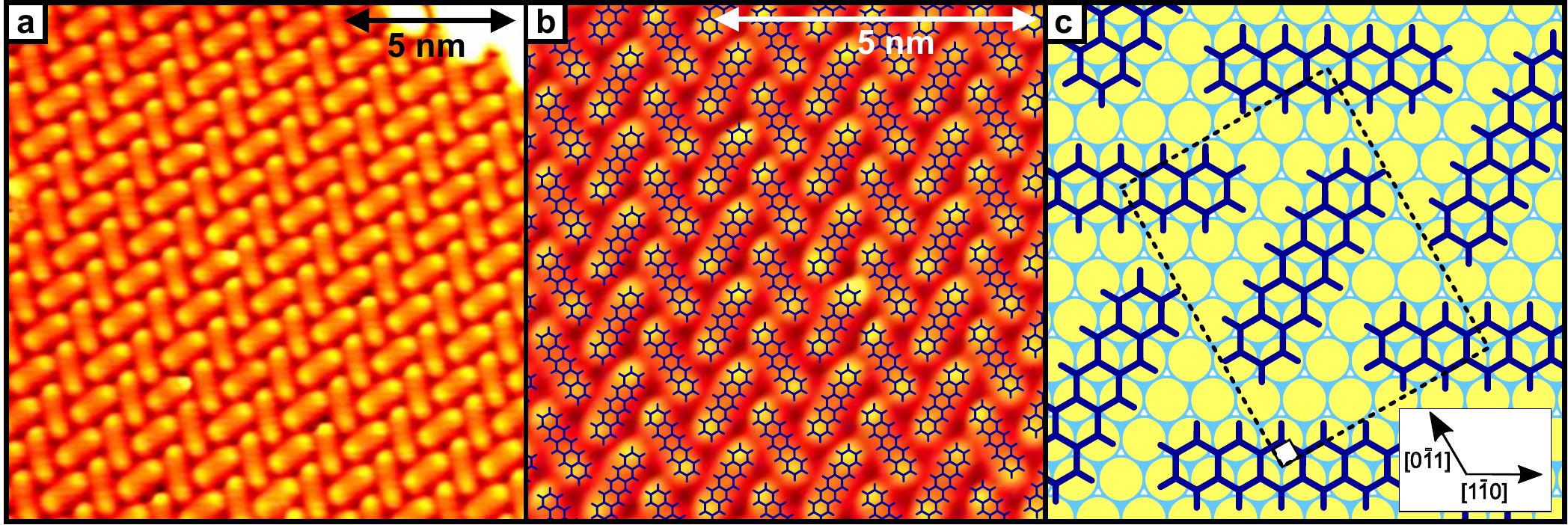}
\caption{\emph{(Color online)} \textbf{(a)}; 15 nm $\times$ 15 nm STM topograph ($V_{gap} = 0.5$ V, $I_T = 0.6$ nA) of Pn/Cu(111), showing the monolayer herringbone structure; \textbf{(b)}; A portion of the image in \emph{(a)} adjusted for drift with the model in \emph{(c)} superimposed; \textbf{(c)};  the proposed Cu(111)-(6 3, 0 7)-2Pn model for herringbone Pn, with Cu substrate vectors for the matrix notation indicated.}
\label{herringbone}
\end{center}
\end{figure*}

Certain molecules in Figure \ref{random} (a) appear \emph{longer} than others.  These are invariably located in larger pores in the film.  The interpretation of this observation is that molecules located in such pores have some freedom of movement.  The elongation of one end of each such molecule, often with a dark feature as well as an additional light feature, rather than an overall loss of resolution over each such site, indicates that the movement of such molecules is activated by the tip.  Tip-induced motion along the long axis of the molecule has been observed by Lagoute \emph{et al.} \cite{Lagoute04-prb}.  The ease of motion along the long axis of the molecule compared to other directions is due to the incommensuration between the periodicity of the acene rings and the Cu NN leading to a potential energy surface with reduced corrugation.

A statistical comparison of apparent molecule `lengths' is shown in Figure \ref{random} (c).  The previous measurement of peak-to-peak Pn length on Cu(111) of $0.98 \pm 0.02$ nm was performed using an LT-STM at 7 K and is directly calibrated from concurrent atomic resolution Cu(111) images \cite{Lagoute04-prb}.  Our VT-STM yields a measurement of $1.07 \pm 0.06$ nm, which is not inconsistent with the previous measurement.  However, we cannot directly observe the Cu(111) surface corrugation.  We thus calibrate our measurement by setting the average length of a (motionless) Pn molecule to the Lagoute value.  The quantity measured in Figure \ref{random} (c) is the \emph{peak separation} of line profiles taken along molecular long axes for molecules in all orientations.  An elongated molecule often has more than one apparent `end', giving rise to more than two peaks in the line profile; these molecules contribute a measurement for each measurable origin--peak distance.  Also, longer-appearing molecules were preferentially sampled.  Therefore, the important feature of the histogram is not the relative population of peaks but the quantization of molecular lengths.  The peaks in the histogram are separated by an average of 0.26 $\pm$ 0.05 nm, which is equal within error to the nearest neighbor (NN) distance in the closely packed rows on the copper surface, 0.2556 nm; thus a similar result to the earlier molecular manipulation study of Pn on Cu(111) \cite{Lagoute04-prb} is obtained.

%new after review
The data under discussion were collected with a range of tunneling parameters. The data presented in Figure \ref{random} were collected at a bias of 0.5 V and tunneling current of 1 nA, which corresponds to a gap impedance of 0.5 G\ohm. Similar molecular motion is present in data collected with gap impedances of 4 G\ohm (2V, 500 pA) and 6 G\ohm (-3V, 500 pA).  These observations contrast with the much lower gap impedance and therefore larger activation energy required for Pn motion in the Lagoute study, in which measurements were conducted with constant tip height on isolated molecules. In this case, molecular motion starts at a tunneling gap impedance of approximately 0.6 M\ohm (0.3V, 500 nA) \cite{Lagoute04-prb}, indicating the tip is much closer to the surface.  It seems, therefore, that the molecular motion depends to a large degree on the local environment of the molecule (i.e., its proximity and orientation relative to its neighbors). An individual molecule on a surface experiences only the attraction to the substrate and therefore resides in a deeper potential well; the intermolecular forces experienced by a molecule in a network such as the one investigated here serve to mitigate this.

\subsubsection{Pn/Cu(111) `2d-herringbone' - $H$-phase}
\begin{figure*}
\begin{center}
\includegraphics[width=0.9\textwidth]{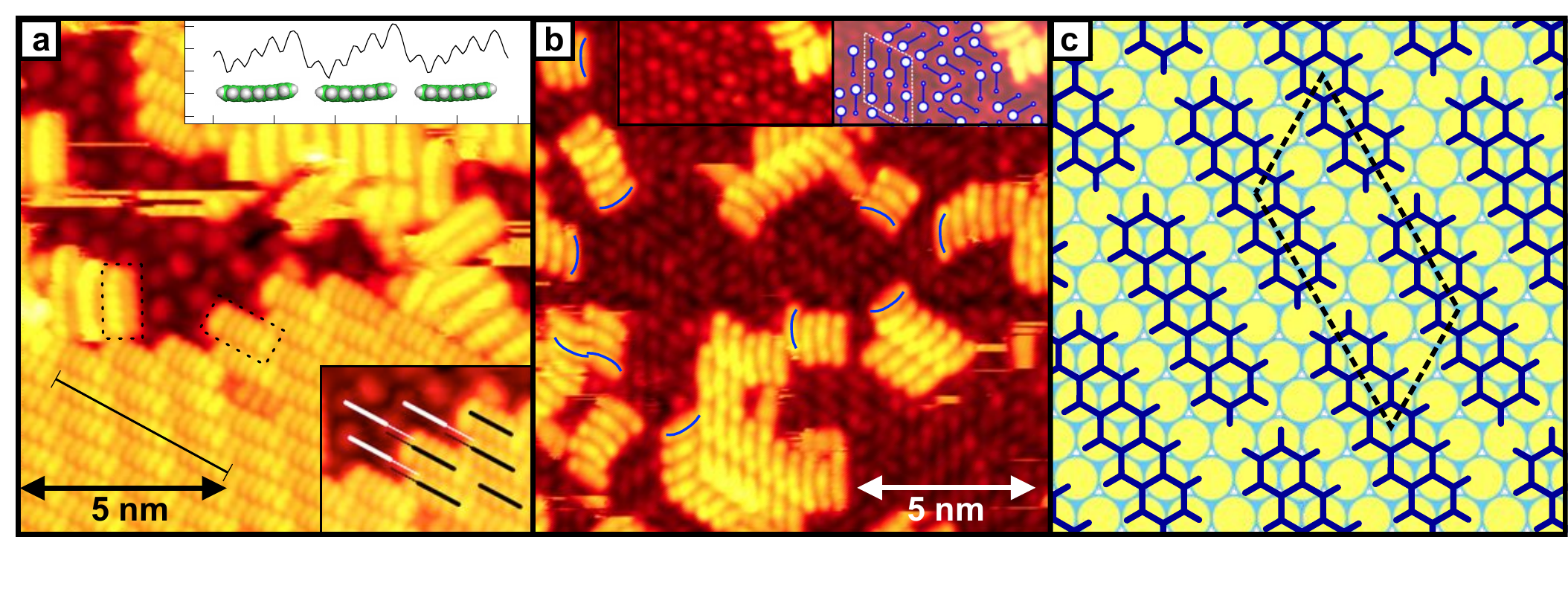}
\caption{\emph{(Color online)} \textbf{(a)}; 12 nm $\times$ 12 nm STM topograph ($V_{gap} = -2.5$ V, $I_T = 0.2$ nA) of Pn/Cu(111), showing both bilayer and monolayer Pn, with the molecules reproduced in Figure \ref{stsfig} indicated with dotted boxes; \emph{(upper inset)} a profile of the line indicated on the bilayer, $x$-scale tick separation is 1 nm, $y$-scale tick separation is 0.02 nm; \emph{(lower inset)} the central portion of the topograph with lines added to highlight the offset between layers; \textbf{(b)};  15 nm $\times$ 15 nm STM topograph ($V_{gap} = 2$ V, $I_T = 0.6$ nA) of Pn/Cu(111), showing bilayer structures of molecules tilted about their molecular long axes, with orientation of crescent-shaped Pn indicated and some monolayer Pn; \emph{(inset)} a section of a topograph at the same scale. The asymmetric blue dumbbells highlight the  tilting of some monolayer molecules about their short axes and a white parallelogram encloses a 6-molecule domain of the high-density monolayer structure given in \emph{(c)}.  The left portion of the inset reproduces the data without overlays for comparison; \textbf{(c)}; the proposed Cu(111)-(3 3, 0 6)-Pn model for the first layer of bilayer Pn - $T$-phase.}
\label{bl}
\end{center}
\end{figure*}

The annealing of a freshly deposited Pn film to 400 K for 10 minutes produces a well ordered monolayer consisting of three rotational domains of large areas of `2d-herringbone' structured material, shown in Figure \ref{herringbone} (a).   Again, knowledge of the adsorption site for a single molecule\cite{Lagoute04-prb} permits the identification of a unique model for the domain structure which reproduces all of the features observed in STM topographs, shown in Figure \ref{herringbone} (b-c).  The matrix notation for the structure observed is Cu(111)-(6 3, 0 7)-2Pn, and the unit cell parameters are $a = 3\sqrt{3}\times 0.2556 = 1.328$~nm, $b = 7\times0.2556 = 1.789 $~nm.  The density of this structure is $\nicefrac{1}{21} = 0.048$ Pn molecules per surface Cu atom.  This density is approximately $\nicefrac{4}{3}$ that of the $R$-phase.  Structural parameters are summarized in Table \ref{table}.  As far as the authors are aware, no similar structure has been observed for Pn or other linear acene on any surface.

\subsubsection{Pn bilayer}

For a film which is only briefly heated to 400 K (as opposed to the 10 minute anneal time required for the herringbone structure), a monolayer packing with higher density than that of the herringbone structure is also observed.  In our data, presented in Figure \ref{bl}(a-b), this structure is only observed in very small domains of at most 20 molecules.  An interesting feature of this monolayer is the development of disorder based on the tilt of the molecules about their short axis.  It is apparent from the STM topograph that the molecular long axes are not uniformly parallel to the substrate (unlike the case for the herringbone structure), with some molecules having one end further from the substrate than the opposite end.  Moreover, `chain-like' behavior occurs, with the higher end of one molecule leading to the lower end of the next molecule.  These phenomena are indicated with the blue dumbbells in the inset to Figure \ref{bl} (b).

Although monolayer domains seem to be limited to small numbers of molecules, \emph{bilayer} domains may grow to a large lateral extent.  The formation of the bilayer apparently stabilizes the monolayer in this structure; if underlying Pn molecules were randomly oriented and/or shifted by a large degree compared to 2nd layer molecules, it would be possible to image them where they protrude from beneath the 2nd layer.  We do not observe such a phenomenon, therefore we conclude that molecules in both layers must largely overlap.  The issue of the registry between the first and second layers will be revisited later in this paper.

The structure for this monolayer is given in Figure \ref{bl} (c).  The matrix notation for this structure is Cu(111)-(3 3, 0 6)-Pn, with Pn molecules oriented along Cu closely packed rows, as for the other structures.  The density of this structure is \nicefrac{1}{18} or 0.057 Pn molecules per surface Cu atom, approximately \nicefrac{5}{3} that of the $R$-phase and \nicefrac{7}{6} that of the $H$-phase.  The density of the monolayer structure giving rise to the bilayer is significantly greater than that of the herringbone structure.  The molecules in the herringbone structure show negligible tilt of the molecules about their molecular short axis.  When the density increases, the steric interaction of the hydrogen atoms leads to the development of a consistent tilt in the molecules. Tilting behavior due to the steric interaction between molecules has been observed also for Pn/Au(111) \cite{Suzuki03-apl} and Pn/Ag-Si(111)-($\sqrt{3}\times\sqrt{3}$)R30$^{\circ}$ \cite{Guaino03-ass,Teng08-ss}, although these systems tilt about the molecular long axis.

\begin{figure}
\begin{center}
\includegraphics[width=0.45\textwidth]{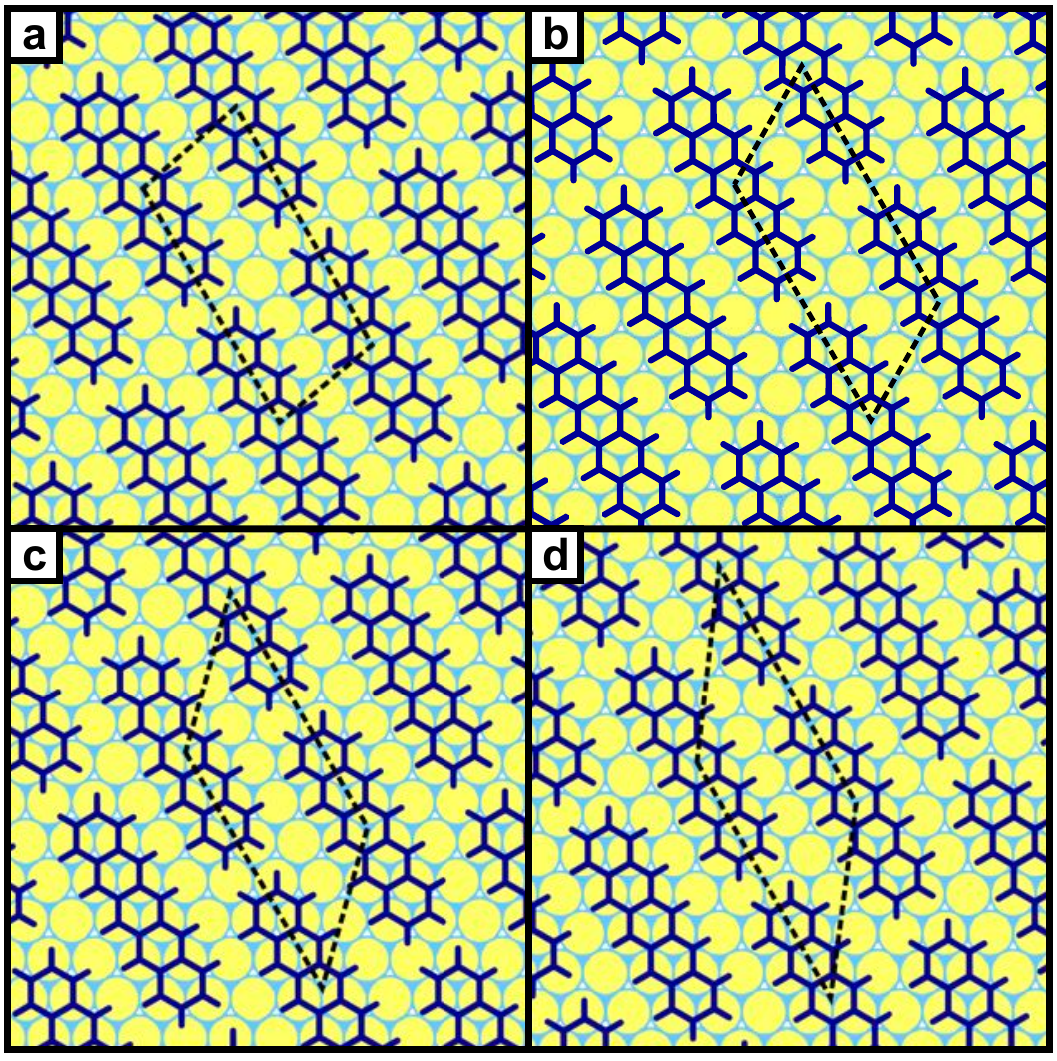}
\caption{\emph{(Color online)}; The 4 non-equivalent registries of Pn end-to-end chains as described in the text.  The mean separation of H nuclei on adjacent molecules differs between structures: \textbf{(a)} 0.271 nm, \textbf{(b)} 0.289 nm, \textbf{(c)} 0.284 nm, \textbf{(d)} 0.286 nm. The maximum separation of 0.289 nm belongs to structure (b) which matches Figure \ref{bl} (c) and is the only one of these structures encountered in the data.}
\label{sterics}
\end{center}
\end{figure}

When quantized to the Cu NN distance, the smallest inter-separation of Pn molecules parallel to their short axes consistently observed across all data is $\nicefrac{3\sqrt{3}}{2}$ Cu$_{NN} = 0.664$ nm and parallel to their long axes is 6 Cu$_{NN} = 1.54$ nm.  If a model is restricted thus and an additional restriction is applied such that molecules should line up end to end (again, as is consistently observed in our data), the only degree of freedom left is the relative positions of end-to-end chains of Pn molecules.  This is also quantized to the Cu NN distance, as the molecular motion data in Figure \ref{random} (c) demonstrate.  Within these restrictions, there exist four non-equivalent registries of adjacent Pn chains.  These are shown in Figure \ref{sterics}.  Only one of these registries (Figure \ref{bl} (c) and Figure \ref{sterics} (b)) is observed in the data (Figure \ref{bl} (a-b)).  By constructing models of small domains of the other 3 registries, a simple analysis measuring the separations between each hydrogen atom on one molecule and its 2 nearest neighbors on adjacent molecules was carried out. We find that the average H - - H nuclear separation for the four structures in Figure \ref{sterics} are (a) 0.271 nm, (b) 0.289 nm, (c) 0.284 nm, (d) 0.286 nm.  The maximum separation, 0.289 nm, belongs to the structure we observe in data and which is shown in Figure \ref{bl} (c).  This suggests that the observed structure is adopted in order to maximize the H - - H nuclear separation and thus minimize the repulsive steric interactions between molecules.

Such bilayers have previously been observed for Pn adsorption both on Au(111) \cite{Kang03-apl} and Ag(111) \cite{Dougherty08-jpcc,Eremtchenko05-prb}.   The Pn multilayer on Au(111) is stabilized via $\pi$-stacking up to the highest coverages observed ($^{\sim}$40 ML).  Kang \emph{et al.} assert that there is no rotational correspondence in the registry of adjacent layers \cite{Kang03-apl}.  The concept that the second layer has dissimilar structure to the monolayer is continued across interpretation of data obtained for Pn/Ag(111), in which Eremtchenko \emph{et al.} conclude that the first layer is disordered \cite{Eremtchenko05-prb}.  However, subsequent studies have somewhat refuted this thesis \cite{Dougherty08-jpcc,Kafer07-cpl} and the similar data for Pn/Cu(111) provide some additional insight into the formation of the bilayer in general as outlined below.

\begin{figure}
\begin{center}
\includegraphics[width=0.45\textwidth]{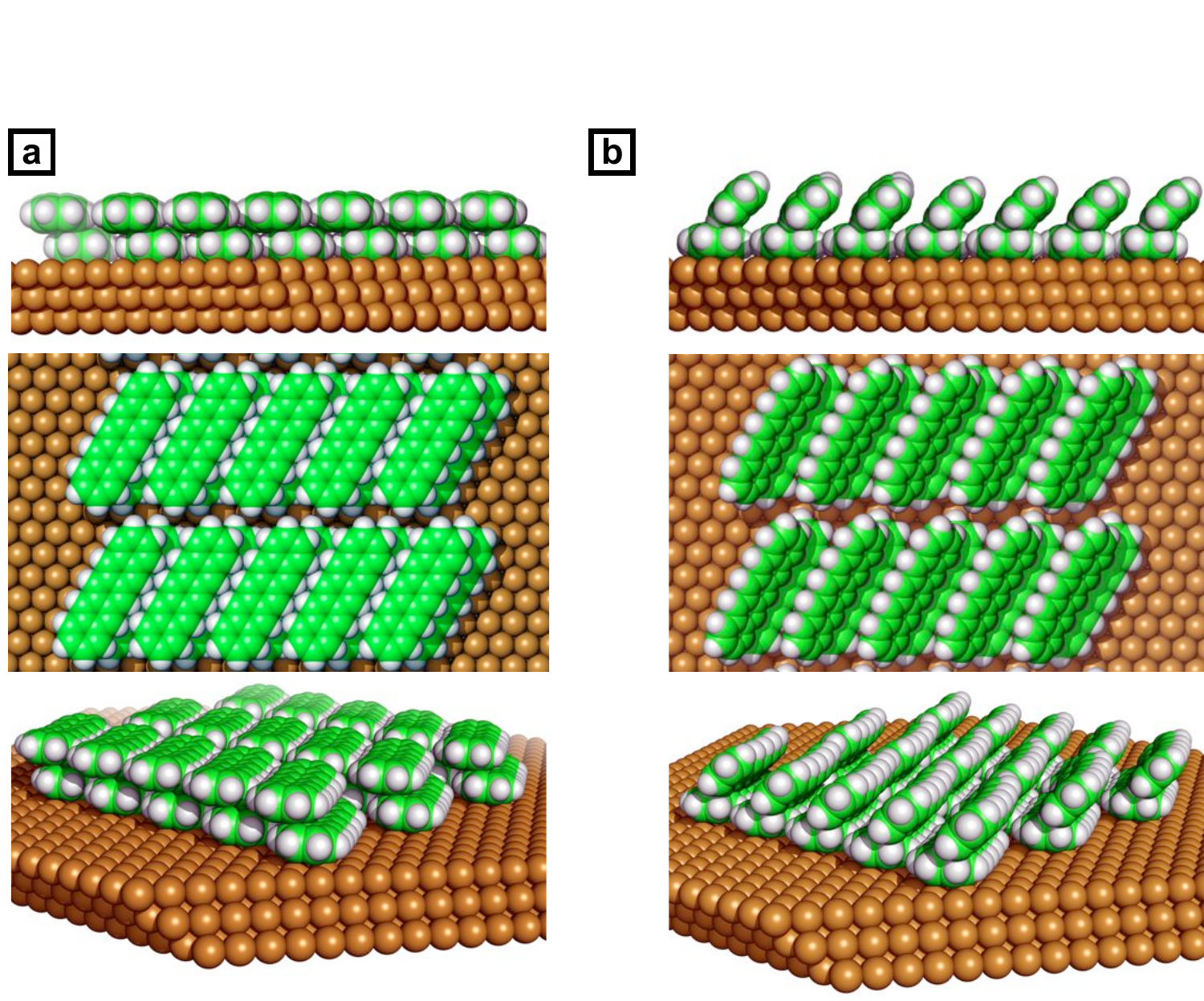}
\caption{\emph{(Color online)} \textbf{(a)}; The proposed model for bilayer Pn with a top layer parallel to the first layer; \textbf{(b)}; The proposed model for bilayer Pn with a tilted top layer.  Both models have as their first layer the structure proposed in Figure \ref{bl}.}
\label{pymolfig}
\end{center}
\end{figure}

%new after review

In the STM topographs in Figure \ref{bl} (a-b), sections of the Pn bilayer adjacent to monolayer regions are shown.  In Figure \ref{bl} (a), recumbent (flat) second layer Pn is  observed in the lower part of the image.  The upper inset in Figure \ref{bl} (a) shows a line profile which highlights a tilt about the molecular short axes for Pn molecules in this layer.  The only possibilities for the origin of such a well-ordered effect are ($i$); the existence of a similar tilt in the molecules beneath; or ($ii$); the existence of an offset between the unit cells in adjacent layers.  Given the subtlety of such a phenomenon and its clear manifestation in the data, either of these possibilities is sufficient to exclude the likelihood of the first layer possessing dissimilar ordering to the second layer.  To further illuminate the structural relationship between the first layer and the second layer, line profiles were extracted over material close to bilayer step edges for those domains that had material in both layers visible.  These line profiles were taken along a Cu close-packed direction at an angle of 60$^\circ$ to the Pn long axis direction and yielded an offset of $0.25 \pm 0.05$ nm for the region investigated; this is shown in the lower inset of Figure \ref{bl}.  This offset was used in the 3D models of the possible bilayer structure depicted in Figure \ref{pymolfig}.  The significance of an offset in this direction is that it results in an oblique volumetric unit cell and thus breaks 2-fold symmetry without the necessity of a tilt about the molecular short axes in the underlying molecules.  Thus it is not possible to assert whether or not molecules in the underlying layer have a tilt about their short axes.

\begin{figure*}
\begin{center}
\includegraphics[width=0.9\textwidth]{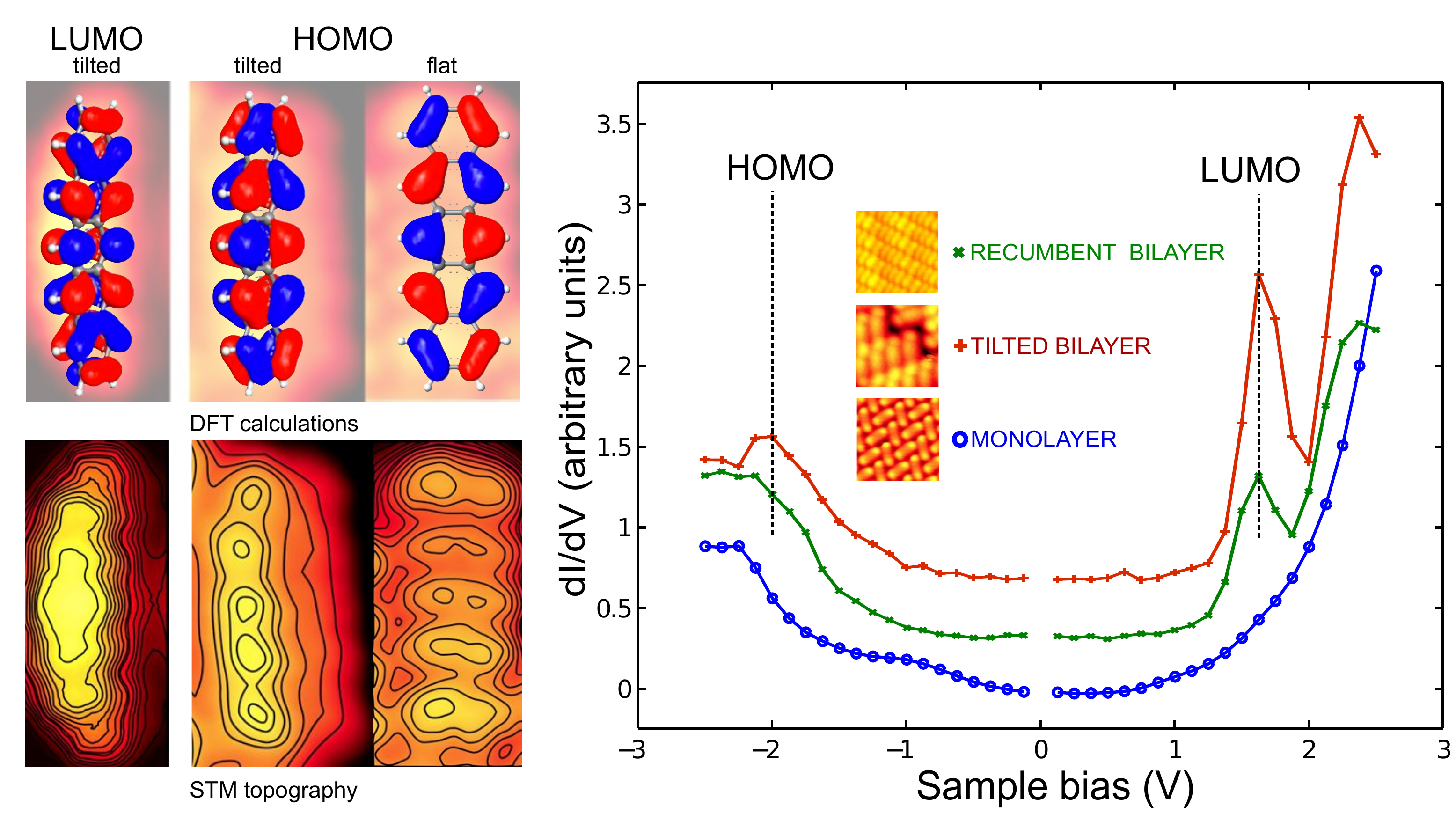}
\caption{\emph{(Color online)} \textbf{Left}; LUMO and HOMO calculations and corresponding STM topographs. There are two HOMO plots as experimental measurements cover Pn molecules in both the recumbent configuration and with a tilt about the molecular long axis.  There is one LUMO plot as LUMO data only cover tilted Pn molecules.  The HOMO images are of individual second layer molecules in the data shown in Figure \ref{bl} (a).  The LUMO image is an average image over several second layer molecules from the data shown in Figure \ref{bl} (b)  in which scans over each molecule are added together, with each scan contributing an equal amount to the overall image; \textbf{Right};  Scanning tunneling spectroscopy (STS) data collected from tilted and recumbent bilayer and monolayer Pn.  The primary orbitals are indicated, showing the lack of contribution to the dI/dV signal of the orbitals of the monolayer.  The STS traces are generated from grid spectroscopy performed over $128\times128$ grids of the surface.   Representative topographs are reproduced for reference. The recumbent 2nd-layer trace is an average of around 11000 curves, the tilted 2nd-layer trace of around 3000 curves, and the monolayer trace is an average of 16000 curves.}
\label{stsfig}
\end{center}
\end{figure*}

Domains of second layer Pn molecules with a tilt about the molecular {\it long} axes are simultaneously observed in the upper portion of Figure \ref{bl} (a) and in Figure \ref{bl} (b).  The distribution of these two kinds of second layer Pn is well represented in the figure; large recumbent bilayer domains are observed, while tilted second-layer Pn is restricted to smaller domains. The attribution of such a tilt to particular domains is based on the greater apparent height of such domains as compared to recumbent second layer Pn and also the crescent shape of Pn molecules in these domains.

It is difficult to understand why molecules in the second layer maintain a bend about their short axis, as shown in Figure \ref{bl} (a).  Chemisorption between the first layer and the Cu substrate involves distortion of the molecules, as found in theory work on Cu(110) \cite{Muller09-prb}.  For molecules adsorbed flat on the surface and imaged with a bias voltage either below the highest occupied molecular orbital (HOMO) or above the lowest unoccupied molecular orbital (LUMO), the shape of the orbitals will affect the apparent shape of the molecules in STM.  However, the side-on observation of crescent-shaped molecules in tilted second layer domains (Figure \ref{bl}(b)) provides convincing evidence that the molecular bending is a genuine geometric effect. The observation of bending in multiple directions for various molecules in Figure \ref{bl} rules out the possibility that these observations are due to an asymmetric tip effect. Such a bend is not visible for first layer Pn molecules adsorbed on Ag or Au surfaces.  Second-layer molecules on such surfaces appear as two-lobed features \cite{Kang03-apl,Dougherty08-jpcc}, though it is not clear whether this is an electronic or geometric effect due to the various tunneling conditions employed.

\begin{table*}
\begin{center}
\begin{tabular}{cccccccccc}
\hline
\multirow{2}{*}{\textbf{System}}      	&  \multirow{2}{*}{\textbf{Matrix notation}	}& \multicolumn{2}{c}{\textbf{Lattice parameter }}(nm) 	& \multirow{2}{*}{\textbf{Density} (\nicefrac{Pn}{Cu})}&	& \multicolumn{2}{c}{\textbf{Electronic states} (eV)}		& \multirow{2}{*}{\textbf{Reference}}	\\
					&						& \textit{expected} 		& \textit{measured}	&							&	& \textit{HOMO}	 			& \textit{LUMO}		&					 	\\
\hline
\rule{0pt}{1.05em}
\multirow{2}{*}{$H$-phase} & \multirow{2}{*}{Cu(111)-(6 3, 0 7)-2Pn}    	& $a = 1.328$			& $a = 1.30\pm0.02$	& \multirow{2}{*}{\nicefrac{1}{21}} 	 		&	& \multirow{2}{*}{--} 			& \multirow{2}{*}{--}&  \multirow{2}{*}{(this work)}		\\ 
			&				& $b = 1.789$ 			& $b = 1.80\pm0.02$	&		&						&			&								&						\\	
\multirow{2}{*}{$T$-phase} &\multirow{2}{*}{Cu(111)-(3 3, 0 6)-2Pn}    	& $a = 0.767$			& $a = 0.78\pm0.02$	& \multirow{2}{*}{\nicefrac{1}{18}} 	 		&	 	Layer 1: 				& --	& -- &\multirow{2}{*}{(this work)}				\\ 
			&				& $b = 1.536$ 			& $b = 1.51\pm0.02$	&								&		Layer 2:				& -2	& 1.6					\\	
$R$-phase & 			& 				& 			& \nicefrac{1}{30} 		&	  & -- & --&(this work)	\\ 

\multirow{2}{*}{Ag(111)-Pn}&&&&&Layer 1:&--&--&&  \\
&&&&&Layer 2:&-2&1.6&\cite{Dougherty08-prb} \\
Au(111)-Pn&&&&&&-0.9&1.3 &  \cite{Soe09-prl}\\
Cu(111)-NaCl-Pn&&&&&&-2.4&1.7&  \cite{Repp05-prl}\\
\hline
\end{tabular}
\end{center}
\caption{Some properties of the structures observed.  Measurements were made on STM topographs with as little as possible drift and are averages over multiple lattice parameters in all observed domain directions.  Once the average was obtained it was calibrated with an average measurement of single Pn molecules from the same image against the earlier calibration for a single molecule \cite{Lagoute04-prb}.}
\label{table}
\end{table*}

The activation of a reorganization of molecules with the STM tip at room temperature has been observed previously for Pn atop Ag-Si(111)-($\sqrt{3}\times\sqrt{3}$)R30$^{\circ}$ \cite{Guaino03-ass,Teng08-ss}.  It is not clear whether a similar phenomenon is responsible for the reorientation of certain molecules in the second layer about their long axis.

K\"{a}fer \emph{et al.} reported that the results for Pn bilayers on Au(111) were not self-consistent \cite{Kafer07-prb}, given that the earlier result of $\pi$-stacked, flat multilayers reported by Kang \emph{et al.} \cite{Kang03-apl} had been obtained after annealing at temperatures ``where sublimation of pentacene already takes place'' \cite{Kafer07-prb}.  The annealing temperature reported in the Kang paper was 353 K, and in the data under discussion ranges from 400-420 K.  They go on to report the observation of bulk Pn(011) islands observed on Au(111), separated by regions of a flat-lying Pn monolayer.  This is used to support a picture of the film in which bulk islands lie on a flat-lying monolayer.  In light of the data reported in this paper showing a recumbent (flat) bilayer and similar data obtained for Pn/Ag(111) \cite{Eremtchenko05-prb,Dougherty08-jpcc}, we conclude that a $\pi$-stacked bilayer or multilayer is a reasonable result.  The additional observation of bilayer material with molecules tilted about their long axes suggests a way that the recumbent structure may relax to the bulk structure.

\subsection{Scanning tunneling spectroscopy}

Although there is great variation in the appearance of the second layer Pn molecules depending on the sample bias during imaging, the monolayer molecules always appear identical, unless, as observed by Lagoute \emph{et al.}, the tip is modified by the adsorption of a Pn molecule \cite{Lagoute04-prb}.  As shown in Figure \ref{stsfig}, bilayer molecules observed above the LUMO energy are 7-lobed whereas those observed below the HOMO energy are 5-lobed.  This observation is in agreement with the calculated wavefunction surfaces shown, where the bend is artificially generated using a spline which maintains bond lengths and reproduces the geometry found using DFT in the case for Pn adsorption atop Cu(110) \cite{Muller09-prb} (which also matches within error the geometry observed for single Pn molecules atop Cu(111)\cite{Lagoute04-prb}).  

The dI/dV signal of the monolayer contains no obvious peaks.  It is possible that the diffuse peak located at -1 V is a hybrid state formed via strong interaction between the Pn HOMO and the $d$-electrons of the substrate. The LUMO peak seems absent altogether for this bias sweep range.

The bilayer, on the other hand, shows extremely clear differential conductance orbital peaks which are located within error precisely at the locations identified by Dougherty \emph{et al.} for a bilayer adsorbed atop Ag(111) \cite{Dougherty08-prb}, and in turn for thick Pn films adsorbed on Au surfaces \cite{Amy05-oe}.  This indicates that, while the first layer molecular orbitals may be strongly affected by the metal $d$-band electrons, the second layer electronic structure is close to that for bulk Pn.  Similar behavior is also noticed for Pn/Cu(119) \cite{Annese08-prb}.  The HOMO-LUMO locations for Pn atop NaCl/Cu(111) are closer to the free-molecule values than to the bulk-Pn values \cite{Repp05-prl}.

In terms of producing good metal-organic contacts suitable for molecular electronics applications, this experiment demonstrates that a Pn monolayer through interaction of its electronic structure with a Cu substrate loses the definition of the HOMO and LUMO states that makes Pn such an attractive candidate for donor action in molecular heterojunctions.  However, the first Pn layer sufficiently isolates the second layer such that it exhibits nearly bulk-like electronic structure. Both of these observations are encouraging for further work based on this system as the donor component of a nanoscale heterojunction device.

\section{Conclusions}

Although Pn has been the subject of intense scrutiny in recent years, two novel structures are observed for monolayer-regime Pn adsorbed on Cu(111).   Neither of these arrangements of Pn molecules have been observed on any substrate.

The development of consistent structural disorder based on molecular tilt about their short axis is also observed, in which increasing density of the layer results in increasing interaction of Pn molecules, ultimately forcing the molecules to twist out of plane about their short axis.  Although this higher density structure is not observed in large monolayer domains, this is the preferred configuration for a bilayer.  This bilayer may occur simultaneously with second layer molecules largely parallel to the substrate and with second layer molecules twisted out of plane about their long molecular axes, in a structure reminiscent of the (011) plane of a bulk Pn crystal \cite{Kafer07-prb,Campbell61-ac}.

Finally, there is emerging a picture of bilayer Pn on multiple metallic substrates in which the first layer is essentially metallic through strong interaction with the substrate and the second layer exhibits molecular orbital energies similar to bulk phase molecules or molecules adsorbed on an insulating layer \cite{Dougherty08-prb,Repp05-prl}.  This is encouraging for energy level alignment in nanoscale heterojunction systems.

\section{Acknowledgments}

The use of the Center for Nanoscale Materials at Argonne National Laboratory was supported by the U.S. Department of Energy, Office of Science, Office of Basic Energy Sciences, under Contract No. DE-AC02-06CH11357.  The authors would like to acknowledge the technical assistance of B. L. Fisher and useful discussions with S. Darling and correspondence with D. Dougherty.  The open-source/freeware software used for analysis and figure editing was Gwyddion (\verb*#gwyddion.net#), PyMOL (\verb*#www.pymol.org#), Inkscape (\verb*#www.inkscape.org#), ImageSXM (\verb*#www.liv.ac.uk/~sdb/ImageSXM/#), ArgusLab (\verb*#www.arguslab.com#) and Jmol (\verb*#jmol.sourceforge.net#).

\end{document}